\documentclass[nofootinbib,twocolumn,showpacs,preprintnumbers,pre,aps]{revtex4-1}

\usepackage{graphicx}
\usepackage{bm}
\usepackage{amsmath}
\usepackage{amssymb}
\usepackage{color}

\begin{document}

\title{Lateral diffusion induced by active proteins in a biomembrane}%

\author{Yuto Hosaka}

\author{Kento Yasuda}

\author{Ryuichi Okamoto}

\author{Shigeyuki Komura}\email{komura@tmu.ac.jp}

\affiliation{
Department of Chemistry, Graduate School of Science and Engineering,
Tokyo Metropolitan University, Tokyo 192-0397, Japan}

\date{April 26, 2017}

\begin{abstract}
We discuss the hydrodynamic collective effects due to active protein molecules that are 
immersed in lipid bilayer membranes and modeled as stochastic force dipoles.
We specifically take into account the presence of the bulk solvent which surrounds the 
two-dimensional fluid membrane. 
Two membrane geometries are considered: the free membrane case and the confined membrane case.
Using the generalized membrane mobility tensors, we estimate the active diffusion coefficient and the 
drift velocity as a function of the size of a diffusing object.
The hydrodynamic screening lengths distinguish the two asymptotic regimes of these quantities.
Furthermore, the competition between the thermal and non-thermal contributions in the total 
diffusion coefficient is characterized by two length scales
corresponding to the two membrane geometries.
These characteristic lengths describe the crossover between different asymptotic behaviors when 
they are larger than the hydrodynamic screening lengths.
\end{abstract}

\maketitle

\section{Introduction}
\label{sec:introduction}

Biomembranes consisting of lipid bilayers can be regarded as thin two-dimensional (2D)
fluids, and membrane protein molecules as well as lipid molecules are allowed to move 
laterally~\cite{Sin72,AlbertsBook}.
These membrane inclusions are subject to the thermal motion of lipid molecules, leading 
to random positional fluctuations.
Such a Brownian motion plays important roles in various life processes such as 
transportation of materials or reaction between chemical species~\cite{LipowskyBook}.
In order to describe lateral diffusion of membrane proteins, a drag coefficient of a cylindrical 
disc moving in a 2D fluid sheet has been theoretically studied in various membrane 
environments~\cite{Saf75,Saf76,Hug81,Evans88,Ramachandran10,Seki11,Seki14}. 
The obtained drag coefficient was used to estimate the diffusion coefficients of membrane 
proteins through Einstein's relation under the assumption that the system is in thermal 
equilibrium~\cite{KomuraBook}.

In recent experiments, however, it has been shown that motions of particles inside cells 
are dominantly driven by random non-thermal forces rather than thermal fluctuations~\cite{Guo14,Par14}.
In these experimental works, they found that non-thermal forces in biological cells are generated 
by active proteins undergoing conformational changes with a supply of adenosine triphosphate (ATP). 
These active fluctuations lead to enhanced diffusion of molecules in the 
cytoplasm~\cite{Yasuda17EPL,Yasuda17PRE}.
Biomembranes also contain various active proteins which, for example, act as ion pumps by 
changing their shapes to exert forces to the adjacent membrane and solvent~\cite{AlbertsBook}.
Lipid bilayers containing such active proteins have been called ``active membranes", 
and their out-of-plane fluctuations (deformations) have already been investigated both 
experimentally and theoretically~\cite{Man99,Ramaswamy00,Man01}. 
However, lateral motions of inclusions  in membranes that are induced by active proteins 
have not yet been considered.
Since such active forces give rise to enhanced diffusion, one needs to take into account 
both active non-thermal fluctuations as well as passive thermal ones to 
calculate diffusion in membranes.

\begin{figure}[tbh]
\centering\includegraphics[scale=0.18]{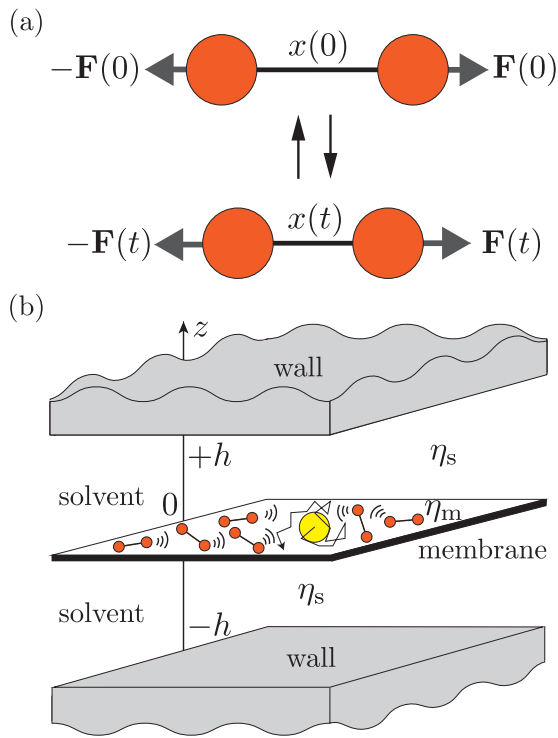}\\
\caption{
(a) The conformational change of an oscillating force dipole representing an active protein.
Within a turnover cycle of the force dipole separated by a distance $x(t)$, it exerts two 
oppositely directed forces $\pm \mathbf{F}(t)$ at time $t$.
The integral intensity of a force dipole is $S$ (see the text). 
(b) Schematic picture showing a flat and infinitely large membrane of 2D viscosity 
$\eta_{\rm m}$ that is located at $z=0$. 
The membrane is surrounded by a bulk solvent of 3D viscosity $\eta_{\rm s}$, and the 
two flat walls are located at $z=\pm h$.
The solvent velocity is assumed to vanish at the surfaces of these walls. 
The ``free membrane" and the ``confined membrane" cases correspond to the limits of 
$h \rightarrow \infty$ and $h \rightarrow 0$, respectively. 
The yellow passive particle undergoes Brownian motion due to thermal and non-thermal
fluctuations. 
The latter contribution is induced by active force dipoles which are homogeneously 
distributed in the membrane with a 2D concentration $c_0$.
}
\label{fig:membrane}
\end{figure}

Recently, Mikhailov and Kapral discussed an enhanced diffusion due to non-thermal 
fluctuating hydrodynamic flows which are induced by oscillating active force dipoles 
[see Fig.~\ref{fig:membrane}(a)]~\cite{Mik15,Kap16}.
They calculated the active diffusion coefficient of a passive particle immersed either 
in a three-dimensional (3D) cytoplasm or in a 2D membrane, and showed that it exhibits 
a logarithmic size dependence for the 2D case.
Moreover, a chemotaxis-like drift of a passive particle was predicted when gradients of 
active proteins or ATP are present~\cite{Mik15}.
Later Koyano \textit{et al.}\ showed that lipid membrane rafts, in which active proteins 
are concentrated, can induce a directed drift velocity near the interface of a domain~\cite{Koy16}.
In these works, they considered membranes that are smaller in size than the hydrodynamic 
screening length.
Huang \textit{et al.}\  performed coarse-grained simulations of active protein inclusions in 
lipid bilayers~\cite{Huang12,Huang13}.
In Ref.~\cite{Huang13}, they showed that active proteins undergoing conformational 
motions not only affect the membrane shape but also laterally stir the lipid bilayer so 
that lipid flows are induced.
Importantly, the flow pattern induced by an immobilized protein resembles the 2D fluid 
velocity fields that are created  by a force dipole.

Following Refs.~\cite{Mik15,Kap16}, we assume that an active protein behaves as an oscillating 
force dipole which acts on the surroundings to generate hydrodynamic flows that can 
induce motions of passive particles in the fluid.
In this paper, we investigate active diffusion and drift velocity of a particle in ``free" 
and ``confined" membranes which are completely flat and infinitely large.
In the free membrane case, a thin 2D fluid sheet is embedded in a 3D solvent having 
typically a lower viscosity than that of the membrane.
Whereas in the confined case, which mimics a supported membrane~\cite{Tan05}, 
a membrane is sandwiched by two rigid walls separated by a finite but small distance 
from it.
For both the free and confined membrane cases, we employ general mobility tensors 
that take into account the hydrodynamic effects mediated by the surrounding 
3D solvent~\cite{Inaura08,Ram11,Ram11b,Komura12}.
Using the general mobility tensors, we numerically calculate the active diffusion 
coefficient and the drift velocity as a function of the diffusing particle size for the 
entire length scales.
Furthermore, several asymptotic expressions are also derived in order to compare 
with numerical estimates and thermal contributions. 
Importantly, our result leads to characteristic length scales describing a crossover 
from non-thermal to thermal diffusive behaviors for large scales.

In the next section, we present the expressions for the active diffusion coefficient 
and the drift velocity in 2D membranes~\cite{Mik15}. 
We also review the general mobility tensors for the free and confined membrane 
cases~\cite{Inaura08,Ram11,Ram11b,Komura12}.
Using these expressions, we calculate in Sec.~\ref{sec:diffusion} the active diffusion 
coefficient for the two geometries. 
In Sec.~\ref{sec:total_diffusion}, we compare the thermal diffusion coefficient with the 
obtained non-thermal diffusion coefficient, and discuss the characteristic crossover lengths.
In Sec.~\ref{sec:drift}, we obtain the drift velocities as a function of the particle size.
The summary of our work and some numerical estimates for the obtained quantities
are given in Sec.~\ref{sec:discussion}.

\section{Active transport and mobility tensors in membranes}
\label{sec:model}

\subsection{Active diffusion coefficient}

Active proteins in a 2D biological membrane, modeled as oscillating force dipoles,
produce non-equilibrium fluctuations and cause an enhancement of the lateral diffusion 
of a passive particle.
We assume that the spatially fixed force dipoles are homogeneously and isotropically 
distributed in the membrane, and they exert only in-plane lateral forces.
The total diffusion coefficient is given by $D=D_{\rm T}+D_{\rm A}$, where $D_{\rm T}$ 
is the thermal contribution and determined by Einstein's relation (which will be discussed 
in Sec.~\ref{sec:total_diffusion}), and $D_{\rm A}$ is the active 
non-thermal contribution given by~\cite{Mik15}
\begin{align}
D_{\rm A}&=\frac{Sc_0}{2}
\int {\rm d}^2r\,
\Omega_{\beta\beta^\prime\gamma\gamma^\prime}
\frac{\partial G_{\alpha\beta}(\mathbf{r})}{\partial r_\gamma}
\frac{\partial G_{\alpha\beta^\prime}(\mathbf{r})}{\partial r_{\gamma^\prime}},
\label{eq:coefficient}
\end{align}
where $\mathbf{r}=(x, y)$ denotes a 2D vector and we have introduced a notation
\begin{align}
\Omega_{\beta\beta^\prime\gamma\gamma^\prime}=
\frac{1}{8} (\delta_{\beta\beta^\prime}\delta_{\gamma\gamma^\prime}+\delta_{\beta\gamma}\delta_{\beta^\prime\gamma^\prime}
+\delta_{\beta\gamma^\prime}\delta_{\beta^\prime\gamma}).
\end{align}
Throughout this paper, the summation over repeated greek indices is assumed.
In Eq.~(\ref{eq:coefficient}), $S$ is the integral intensity of a force dipole, $c_0$ is 
the constant 2D concentration of active proteins, and $G_{\alpha\beta}(\mathbf{r})$ 
is the membrane mobility tensor which will be discussed later separately.

Within a fluctuating ``dimer model" as presented in Fig.~\ref{fig:membrane}(a), 
the magnitude of a force dipole is given by $m(t)=x(t)F(t)$, where $x(t)$ is the 
distance between the two spheres and $F(t)$ is the magnitude of the oppositely 
directed forces.
The statistical average of the dipole magnitude vanishes, i.e., 
$\langle m(t)\rangle=0$, whereas the integral intensity $S$ of a force dipole is given by 
$S=\int_0^\infty {\rm d}t\, \langle m(t)m(0)\rangle$~\cite{Mik15}.
Since we assume that active proteins are homogeneously distributed in the membrane 
as shown in Fig.~\ref{fig:membrane}(b), it is sufficient to consider only the isotropic 
diffusion as given by Eq.~(\ref{eq:coefficient}).

In deriving Eq.~(\ref{eq:coefficient}), the size of a dipole is assumed to be much smaller 
than the distance between the passive particle and active force dipoles~\cite{Mik15}. 
At large distances, almost any object that changes its shape would create a flow field
that corresponds to some force dipole. 
It should be noted, however,  that the above expression is not accurate when the distance 
between them becomes smaller. 
As for the mobility tensor in 3D fluids, it is known that the Rotne-Prager mobility tensor
takes into account higher order corrections to the Oseen mobility tensor and gives more 
accurate approximation at short distances~\cite{Kap16}.
Such a better approximation has not been worked out so far for 2D fluid membranes, 
and we shall only consider the lowest order contribution (see later calulations).
In the above, we have also assumed that force dipoles are spatially fixed in the membrane. 
Since no forces are applied to fix the dipoles, such an approximation is justified when 
the dynamics of force dipoles is much slower than that of the passive particle.

\subsection{Drift velocity}

Although we have assumed above that $c_0$ is constant, 
active proteins are often distributed inhomogeneously in the membrane due to 
heterogeneous structures such as sphingolipid-enriched domains~\cite{Sim97,Kom14}.
According to the ``lipid raft" hypothesis, theses domains act as platforms for membrane 
signaling and trafficking~\cite{Lin10}.
Hence it is also important to consider the effects of nonuniform spatial distribution of 
active proteins and to see how it affects the lateral dynamics in membranes.

When a spatial concentration gradient $\nabla c$ of active protein is present, it gives 
rise to an unbalanced induced forces between two points in the membrane.
Hence passive particles are subjected to a drift toward either lower or higher concentration 
of active proteins, and a chemotaxis-like drift can occur. 
When the absolute value of the concentration gradient $\vert \nabla c \vert$ is assumed 
to be constant, the induced drift velocity of a passive particle in the direction  
$\nabla c$ is given by~\cite{Mik15}
\begin{align}
V&=-S \vert \nabla c  \vert  
\int {\rm d}^2r\,
\Omega_{\beta\beta^\prime\gamma\gamma^\prime}
\hat{n}_\alpha\frac{\partial^2 G_{\alpha\beta}(\mathbf{r})}{\partial r_\gamma\partial r_\delta}\frac{\partial G_{\delta\beta^\prime}(\mathbf{r})}{\partial r_{\gamma^\prime}}(\mathbf{r}\cdot\hat{\mathbf{n}}).
\label{eq:drift}
\end{align}
Here, the unit vector $\hat{\mathbf{n}}=\nabla c/\vert \nabla c \vert$ denotes the direction 
of the concentration gradient of active proteins.
We shall employ the above expression to obtain the lateral drift velocity in a membrane
by using the membrane mobility tensor as discussed below.

\subsection{Membrane mobility tensors}

Since we discuss active diffusion in an infinitely large flat membrane, we use the 2D membrane 
mobility tensor which also takes into account the hydrodynamic effects of the surrounding 3D 
solvent.
We consider a general situation as depicted in Fig.~\ref{fig:membrane}(b), where a fluid 
membrane of 2D shear viscosity $\eta_{\rm m}$ is surrounded by a solvent of 3D 
shear viscosity $\eta_{\rm s}$. 
Furthermore, we consider the case in which there are two walls located symmetrically 
at an arbitrary distance $h$ from the flat membrane~\cite{Inaura08,Ram11,Ram11b,Komura12}.

We denote the in-plane velocity vector of the fluid membrane by ${\mathbf v}({\mathbf r})$
and the lateral pressure by $p({\mathbf r})$.
Assuming that the incompressibility condition holds for the fluid membrane, we write its 
hydrodynamic equations as
\begin{align}
&\nabla \cdot {\mathbf v} = 0, \\
& \eta_{\rm m} \nabla^2 {\mathbf v} - \nabla p + {\mathbf f}_{\rm s} + {\mathbf F}=0.
\end{align}
The second equation is the 2D Stokes equation, where ${\mathbf f}_{\rm s}$ is the force 
exerted on the membrane by the surrounding solvent, and ${\mathbf F}$ is any external 
force acting on the membrane.
If we denote the upper and lower solvents with the superscripts $\pm$,
the two solvent velocities ${\mathbf v}^{\pm}({\mathbf r},z)$ and
pressures $p^{\pm}({\mathbf r},z)$ obey the following hydrodynamic equations,
respectively
\begin{align}
& \widehat{\nabla} \cdot {\mathbf v}^{\pm} = 0, \\
& \eta_{\rm s} \widehat{\nabla}^2 {\mathbf v}^{\pm} - \widehat{\nabla} p^{\pm}= 0,
\end{align}
where $\widehat{\nabla}$ stands for the 3D differential operator.

We assume that the surrounding solvent cannot permeate the membrane, 
and impose the no-slip boundary condition between the membrane and the surrounding solvent
at $z=0$~\cite{Saf75,Saf76,Inaura08,Ram11,Ram11b,Komura12}.
Hence we require the conditions 
\begin{equation}
v_z^{\pm}({\mathbf r},0)=0,~~~~~
v_{\alpha}({\mathbf r})=v_{\alpha}^{\pm}({\mathbf r},0),
\end{equation}
where $\alpha=x, y$.
Furthermore, the solvent velocity vanishes at the walls located at $z=\pm h$, i.e., 
$v_{\alpha}^{\pm}({\mathbf r},\pm h)=0$.

By solving the above coupled hydrodynamic equations in Fourier space 
with $\mathbf{k}=(k_x,k_y)$ being the 2D wavevector, the 2D mobility
tensor ${G}_{\alpha\beta}({\mathbf k})$ defined through
${ v_\alpha}({\mathbf k})= G_{\alpha\beta}({\mathbf k})  { F}_{\beta}({\mathbf k})$
can be obtained as~\cite{Inaura08,Ram11,Ram11b,Komura12}
\begin{align}
G_{\alpha\beta}(\mathbf{k})=\frac{\delta_{\alpha\beta}-\hat{k}_\alpha \hat{k}_\beta}
{\eta_{\rm m}\left[k^2+\nu k\coth(kh)\right]},
\label{eq:general_mobility}
\end{align}
where $k= |\mathbf{k}|$ and $\hat{k}_\alpha=k_\alpha/k$, and 
the ratio of the two viscosities $\nu^{-1}=\eta_{\rm m}/(2\eta_{\rm s})$ defines the 
Saffman-Delbr\"{u}ck (SD) hydrodynamic screening length~\cite{Saf75,Saf76}.
Notice that $\eta_{\rm m}$ and $\eta_{\rm s}$ have different dimensions, 
and $\nu^{-1}$ has a dimension of length.

In order to perform analytical calculations, the two limiting cases of 
Eq.~(\ref{eq:general_mobility}) are considered, i.e., 
the ``free membrane" case and the ``confined membrane" case corresponding 
to the limits of $h \rightarrow \infty$ and $h \rightarrow 0$, respectively~\cite{Ram11,Ram11b,Komura12}.    
For the free membrane case, we take the limit $kh\gg1$ in Eq.~(\ref{eq:general_mobility})
and obtain the following asymptotic expression
\begin{align}
G^{\rm F}_{\alpha\beta}(\mathbf{k})&=\frac{\delta_{\alpha\beta}-\hat{k}_\alpha \hat{k}_\beta}
{\eta_{\rm m}(k^2+\nu k)}.
\label{eq:ft_sd}
\end{align}
Hereafter, we shall denote the quantities for the free membrane case with the 
superscript ``F''.
For the confined membrane case, on the other hand, we take the opposite limit 
$kh \ll 1$ and obtain 
\begin{align}
G^{\rm C}_{\alpha\beta}(\mathbf{k})&=\frac{\delta_{\alpha\beta}-\hat{k}_\alpha \hat{k}_\beta}
{\eta_{\rm m}(k^2+\kappa^2)},
\label{eq:ft_es}
\end{align}
where $\kappa^{-1}=(h/\nu)^{1/2}$ is the Evans-Sackmann (ES) screening 
length~\cite{Evans88}, and we use the superscript ``C'' for the quantities related to the 
confined membrane case. 
We note that the ES screening length $\kappa^{-1}$ is the geometric mean of 
$\nu^{-1}$ and $h$ so that we typically have $\kappa^{-1} < \nu^{-1}$.

Taking the inverse Fourier transform of Eqs.~(\ref{eq:ft_sd}) and (\ref{eq:ft_es}), 
we obtain the mobility tensors in the real space for the two limiting cases as~\cite{Ram11,Ram11b,Komura12} 
\begin{align}
G_{\alpha\beta}^{{\rm F}}(\mathbf{r})=&\frac{1}{4\eta_{\rm m}}
\left[\mathbf{H}_0(\nu r)-Y_0(\nu r)+\frac{2}{\pi\nu^2 r^2}\right.\nonumber \\
&\left.-\frac{\mathbf{H}_1(\nu r)}{\nu r}
+\frac{Y_1(\nu r)}{\nu r}
\right]\delta_{\alpha\beta}\notag\\
&+\frac{1}{4\eta_{\rm m}}\left[
-\frac{4}{\pi\nu^2 r^2}+\frac{2\mathbf{H}_1(\nu r)}{\nu r}\notag\right.\\
&\left.-\frac{2Y_1(\nu r)}{\nu r}-\mathbf{H}_0(\nu r)+Y_0(\nu r)
\right]\hat{r}_\alpha\hat{r}_\beta,
\label{eq:mobility_sd}
\end{align}
and
\begin{align}
G_{\alpha\beta}^{{\rm C}}(\mathbf{r})=&\frac{1}{2\pi\eta_{\rm m}}\left[
K_0(\kappa r)+\frac{K_1(\kappa r)}{\kappa r}-\frac{1}{\kappa^2r^2}
\right]\delta_{\alpha\beta}\notag\\
&+\frac{1}{2\pi\eta_{\rm m}}\left[
-K_0(\kappa r)-\frac{2K_1(\kappa r)}{\kappa r}+\frac{2}{\kappa^2r^2}
\right]\hat{r}_\alpha\hat{r}_\beta,
\label{eq:mobility_es}
\end{align}
respectively, where we have used the notations 
$r= |\mathbf{r}|$ and $\hat{r}_\alpha=r_\alpha/r$.
In the above, $\mathbf{H}_n(z)$ are the Struve functions, $Y_n(z)$ the Bessel functions 
of the second kind, and $K_n(z)$ the modified Bessel functions of the second kind.
The physical meaning of the above expressions was also discussed in 
Refs.~\cite{Dia09,Opp09,Opp10}.
We note that if there is only one wall instead of two walls, the definition of the ES length 
needs to be modified as $\kappa^{-1} \rightarrow (2h/\nu)^{1/2}$~\cite{Opp10}.
In the next sections, we shall use Eqs.~(\ref{eq:mobility_sd}) and (\ref{eq:mobility_es}) to 
calculate the active diffusion coefficients and the drift velocity.

\section{Active diffusion coefficient}
\label{sec:diffusion}

\subsection{Free membranes}

We first calculate the active diffusion coefficient for the free membrane case by substituting Eq.~(\ref{eq:mobility_sd}) into Eq.~(\ref{eq:coefficient}). 
Since the integrand in Eq.~(\ref{eq:coefficient}) diverges logarithmically at short distances, 
we need to introduce a small cutoff length $\ell_{\rm c}$.
Physically, $\ell_{\rm c}$ is given by the sum of the size of a passive particle 
(undergoing lateral Brownian motion) and that of a force dipole~\cite{Kap16}.
In the following, we generally assume that force dipoles are smaller than the diffusing object whose
size is represented by $\ell_{\rm c}$. 
This is further justified when we consider lateral diffusion of a passive object that is larger than 
the SD or ES screening lengths.

\begin{figure}[tbh]
\begin{center}
\includegraphics[scale=0.35]{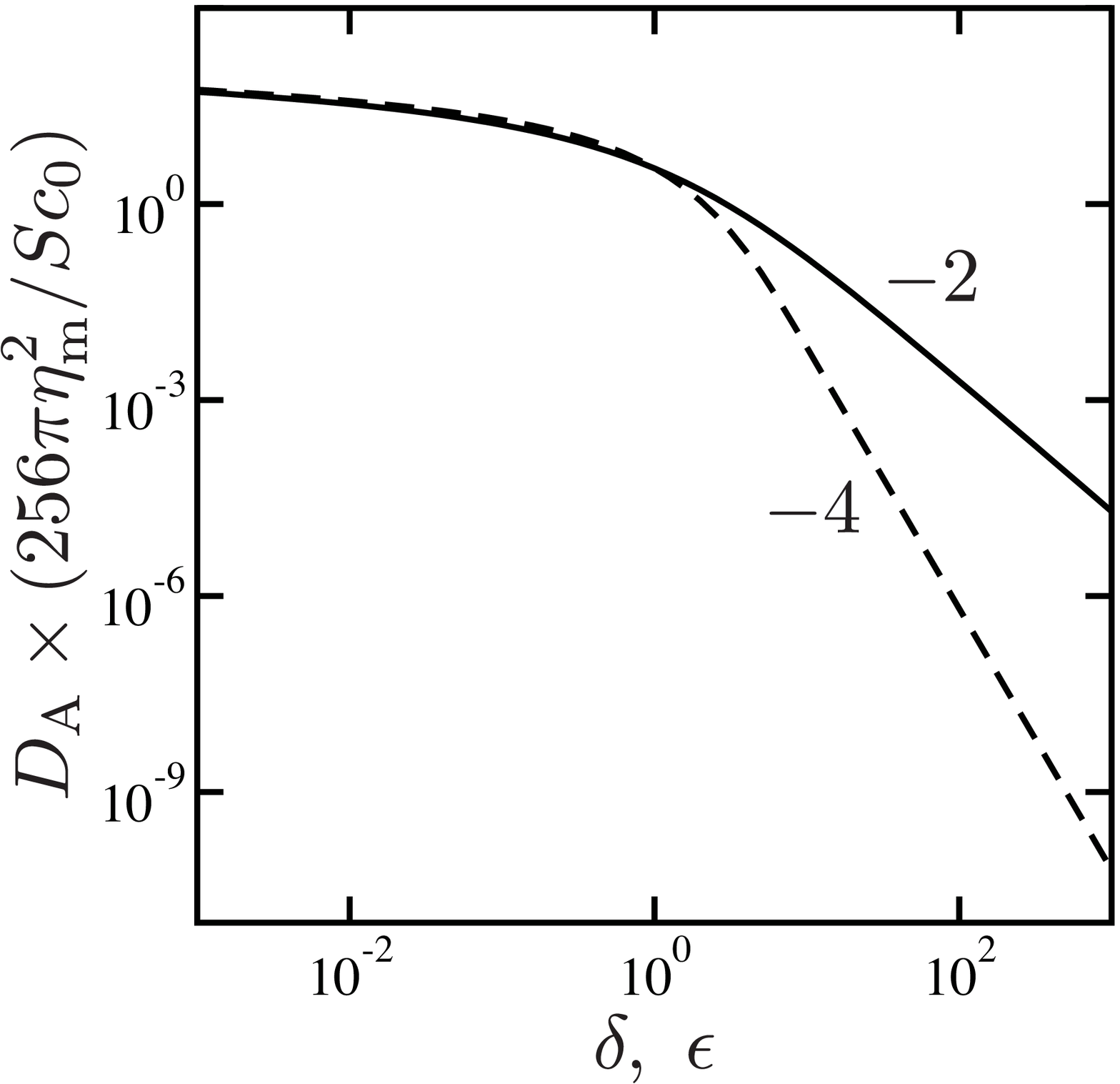}
\end{center}
\caption{
The plot of the scaled active diffusion coefficient $D_{\rm A}$ as a function 
of the scaled cutoff length $\delta=\nu\ell_{\rm c}$ and $\epsilon = \kappa\ell_{\rm c}$ 
for the free membrane case [solid line, see Eq.~(\ref{eq:diffusion_sd})] and the confined 
membrane case [dashed line, see Eq.~(\ref{eq:diffusion_es})], respectively.
Here $D_{\rm A}$ is scaled by $Sc_0/(256\pi\eta_{\rm m}^2)$. 
The numbers in this plot indicate the slope of the curves and represent the powers 
of the algebraic dependencies.
}
\label{fig:sd_es}
\end{figure}

Introducing a dimensionless vector $\mathbf{z}=\nu\mathbf{r}$ scaled by the 
SD length, we can write the active diffusion coefficient for the free membrane case as
\begin{align}
D^{{\rm F}}_{\rm A}=&\frac{Sc_0}{32\pi^2\eta_{\rm m}^2}
\int_{\delta}^\infty
{\rm d}^2z\,\Omega_{\beta\beta^\prime\gamma\gamma^\prime}
\frac{\partial g^{{\rm F}}_{\alpha\beta}(\mathbf{z})}{\partial z_\gamma}\frac{\partial g^{{\rm F}}_{\alpha\beta^\prime}(\mathbf{z})}{\partial z_{\gamma^\prime}},
\label{eq:diffusion_sd}
\end{align}
where $\delta =\nu \ell_{\rm c}$ is the dimensionless cutoff, and
$g^{{\rm F}}_{\alpha\beta}(\mathbf{z})=4\pi\eta_{\rm m}G^{\rm F}_{\alpha\beta}$ 
is the corresponding dimensionless mobility tensor [see Eq.~(\ref{eq:mobility_sd})].
We have first evaluated the above integral numerically. 
In Fig.~\ref{fig:sd_es}, we plot the obtained $D^{{\rm F}}_{\rm A}$ as a function of 
$\delta=\nu \ell_{\rm c}$ by the solid line. 
We see that the active diffusion coefficient depends only weakly on the particle size at small scales, 
whereas it shows a stronger size dependence described by a power-law behavior at large scales.
The crossover between these two behaviors is set by the condition $\delta\approx1$.

In order to understand the above behaviors, we next discuss the asymptotic behaviors of 
$D^{{\rm F}}_{\rm A}$ for both small and large $\delta$ values.
Expanding the mobility tensor in Eq.~(\ref{eq:mobility_sd}) for $\nu r \ll 1$ and $\nu r \gg 1$, 
we have~\cite{Opp10}
\begin{align}
g_{\alpha\beta}^{{\rm F}}(\mathbf{z})&\approx\left(\ln\frac{2}{z}-\gamma-\frac{1}{2}\right)\delta_{\alpha\beta}
+\hat{z}_\alpha\hat{z}_\beta,
\label{eq:mobility_sd_small}
\end{align}
and 
\begin{align}
g_{\alpha\beta}^{{\rm F}}(\mathbf{z})&\approx\frac{2}{z}\hat{z}_\alpha\hat{z}_\beta,
\label{eq:mobility_sd_large}
\end{align}
respectively, where $\gamma=0.5722\cdots$ is Euler's constant.
By substituting Eqs.~(\ref{eq:mobility_sd_small}) and (\ref{eq:mobility_sd_large}) into Eq.~(\ref{eq:diffusion_sd}), we can analytically obtain the asymptotic forms of the active 
diffusion coefficient as a function of $\delta =\nu \ell_{\rm c}$.

As obtained in Ref.~\cite{Mik15}, we find for $\delta\ll1$
\begin{align}
D^{{\rm F}}_{\rm A}&\approx\frac{Sc_0}{32\pi\eta_{\rm m}^2}
\ln \frac{L}{\ell_{\rm c}},
\label{eq:diffusion_sd_small}
\end{align}
where a large cutoff length $L$ is introduced because the integral in 
Eq.~(\ref{eq:diffusion_sd}) also diverges logarithmically at large distances.
In order to match with the numerical estimation, we obtain $L\approx0.682 \nu^{-1}$.
The above logarithmic dependence on $\ell_{\rm c}$ means that $D^{{\rm F}}_{\rm A}$
depends only weakly on the particle size. 
We also note that the above expression contains only the membrane viscosity 
$\eta_{\rm m}$, and does not depend on the solvent viscosity $\eta_{\rm s}$.
This is because the hydrodynamics at small scales is primarily dominated by the 2D 
membrane property.

In the opposite limit of $\delta  \gg1$, on the other hand, we show in the 
Appendix A that the active diffusion coefficient becomes
\begin{align}
D^{{\rm F}}_{\rm A}&\approx\frac{5Sc_0}{256\pi\eta_{\rm s}^2}
\frac{1}{\ell_{\rm c}^2},
\label{eq:diffusion_sd_large}
\end{align}
which is an important result of this paper.
This asymptotic expression decays as $1/\ell_{\rm c}^2$ and depends now only 
on $\eta_{\rm s}$, indicating that the membrane lateral dynamics is governed by 
the surrounding 3D fluid at large scales.
From the obtained asymptotic expressions in Eqs.~(\ref{eq:diffusion_sd_small}) and 
(\ref{eq:diffusion_sd_large}), the behavior of $D^{{\rm F}}_{\rm A}$ in 
Fig.~\ref{fig:sd_es} is explained as a crossover from a logarithmic dependence to 
an algebraic dependence with a power of $-2$.

\subsection{Confined membranes}

Next we consider the confined membrane case.
With the use of Eq.~(\ref{eq:mobility_es}) the active diffusion coefficient can be written as 
\begin{align}
D^{\rm C}_{\rm A}&=\frac{Sc_0}{32\pi^2\eta_{\rm m}^2}
\int_{\epsilon}^\infty
{\rm d}^2w\,\Omega_{\beta\beta^\prime\gamma\gamma^\prime}
\frac{\partial g^{{\rm C}}_{\alpha\beta}(\mathbf{w})}{\partial w_\gamma}\frac{\partial g^{{\rm C}}_{\alpha\beta^\prime}(\mathbf{w})}{\partial w_{\gamma^\prime}},
\label{eq:diffusion_es}
\end{align}
where $\mathbf{w}=\kappa\mathbf{r}$ is a different dimensionless variable, 
$\epsilon =\kappa \ell_{\rm c}$ is a differently scaled cutoff, and 
$g^{{\rm C}}_{\alpha\beta}(\mathbf{w})=4\pi\eta_{\rm m}G^{\rm C}_{\alpha\beta}$
is the corresponding dimensionless mobility tensor [see Eq.~(\ref{eq:mobility_es})].
Performing the numerical integration of Eq.~(\ref{eq:diffusion_es}), we plot 
in Fig.~\ref{fig:sd_es} the active diffusion coefficient $D^{\rm C}_{\rm A}$ as a function 
of $\epsilon=\kappa \ell_{\rm c}$ by the dashed line. 
For small $\epsilon$ values, the behavior of $D^{\rm C}_{\rm A}$ is similar to that of 
$D^{\rm F}_{\rm A}$, while $D^{\rm C}_{\rm A}$ decays much faster than 
$D^{\rm F}_{\rm A}$ for large $\epsilon$ values.

To discuss these size dependencies, we use the asymptotic expressions of 
Eq.~(\ref{eq:mobility_es}) for $\kappa r \ll 1$ and $\kappa r \gg 1$ given by~\cite{Opp10} 
\begin{align}
g_{\alpha\beta}^{{\rm C}}(\mathbf{w})&\approx\left(\ln\frac{2}{w}-\gamma-\frac{1}{2}\right)\delta_{\alpha\beta}
+\hat{w}_\alpha\hat{w}_\beta,
\label{eq:mobility_es_small}
\end{align}
and
\begin{align}
g_{\alpha\beta}^{{\rm C}}(\mathbf{w})&\approx-\frac{2}{w^2}
(\delta_{\alpha\beta}-2\hat{w}_\alpha\hat{w}_\beta),
\label{eq:mobility_es_large}
\end{align}
respectively.
Note that Eq.~(\ref{eq:mobility_es_small}) is identical to Eq.~(\ref{eq:mobility_sd_small}) when 
$w$ is replaced by $z$.
Hence, in the limit of $\epsilon\ll1$, the active diffusion coefficient for the confined membrane 
case should be identical to Eq.~(\ref{eq:diffusion_sd_small}) and is given by~\cite{Mik15} 
\begin{align}
D^{\rm C}_{\rm A}&\approx\frac{Sc_0}{32\pi\eta_{\rm m}^2}
\ln\frac{L}{\ell_{\rm c}}.
\label{eq:diffusion_es_small}
\end{align}
The large cutoff length should be taken here as $L \approx 1.12 \kappa^{-1}$. 
As mentioned before, the 2D hydrodynamic effect is more important at small scales,
and $D^{\rm C}_{\rm A}$ is logarithmically dependent on the particle size.

In the large size limit of $\epsilon\gg1$, on the other hand, we also show in the 
Appendix A that $D^{\rm C}_{\rm A}$ asymptotically behaves as 
\begin{align}
D^{\rm C}_{\rm A}&\approx\frac{Sc_0}{16\pi\eta_{\rm s}^2}
\frac{h^2}{\ell_{\rm c}^4},
\label{eq:diffusion_es_large}
\end{align}
which is another important result.
The obtained expression decays as $1/\ell_{\rm c}^4$ which is much stronger 
than Eq.~(\ref{eq:diffusion_sd_large}) for the free membrane case. 
According to Eqs.~(\ref{eq:diffusion_es_small}) and (\ref{eq:diffusion_es_large}),
the behavior of $D^{{\rm C}}_{\rm A}$ in Fig.~\ref{fig:sd_es} can be understood 
as a crossover from a logarithmic dependence to an algebraic dependence with a 
power of $-4$.

\section{Total diffusion coefficient}
\label{sec:total_diffusion}

Having obtained the active diffusion coefficients for the free and the confined
membrane cases, we now discuss the total lateral diffusion coefficients in membranes 
by considering both thermal and non-thermal contributions. 
Concerning the thermal diffusion coefficient $D^{\rm F}_{\rm T}$ for the free 
membrane case, we use an empirical expression obtained by Petrov and 
Schwille~\cite{Pet08,Petrov12}
\begin{align}
D^{\rm F}_{\rm T}(\delta)=&\frac{k_{\rm B}T}{4\pi\eta_{\rm m}}\left[\ln\frac{2}{\delta}-\gamma+\frac{4\delta}{\pi}-\frac{\delta^2}{2}\ln\frac{2}{\delta}\right]\nonumber \\
& \times \left[1-\frac{\delta^3}{\pi}\ln\frac{2}{\delta}+\frac{c_1\delta^{b_1}}{1+c_2\delta^{b_2}}\right]^{-1},
\label{DFT}
\end{align}
where $k_{\rm B}$ is the Boltzmann constant, $T$ is the temperature, and the four numerical 
constants are chosen as $c_1=0.73761$, $b_1=2.74819$, $c_2=0.52119$, 
and $b_2=0.51465$~\cite{Petrov12}.
For the free membrane case, there is no exact analytical expression of the thermal diffusion 
coefficient which covers the entire size range, except for the case where a 2D polymer chain 
is confined in a fluid membrane~\cite{Ram11}.
Equation~(\ref{DFT}) is known to recover the correct asymptotic limits of the thermal diffusion 
coefficients both for $\delta \ll 1$~\cite{Saf75,Saf76} and $\delta \gg 1$~\cite{Hug81}.

\begin{figure}[tbh]
\begin{center}
\includegraphics[scale=0.35]{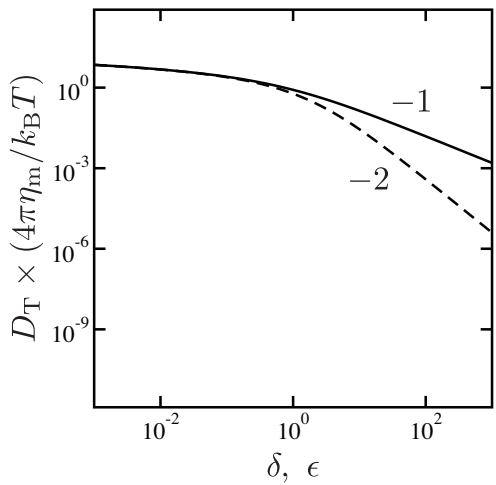}
\end{center}
\caption{
The plot of the scaled thermal diffusion coefficient $D_{\rm T}$ as a function 
of the scaled cutoff length $\delta=\nu\ell_{\rm c}$ and $\epsilon = \kappa\ell_{\rm c}$ 
for the free membrane case [solid line, see Eq.~(\ref{DFT})] and the confined 
membrane case [dashed line, see Eq.~(\ref{DCT})], respectively.
Here $D_{\rm T}$ is scaled by $k_{\rm B}T/(4\pi\eta_{\rm m})$. 
The numbers in this plot indicate the slope of the curves and represent the powers 
of the algebraic dependencies. 
}
\label{fig:thermal_sd_es}
\end{figure}

On the other hand, the thermal diffusion coefficient $D^{\rm C}_{\rm T}$ for the confined 
membrane case was explicitly calculated by Evans {\it et al.}~\cite{Evans88} and also by
Ramachandran \textit{et al.}~\cite{Ramachandran10,Seki11,Seki14,KomuraBook}.
In this case, the resulting expression is given by 
\begin{align}
D^{\rm C}_{\rm T}(\epsilon)=&\frac{k_{\rm B}T}{4\pi\eta_{\rm m}}\left[\frac{\epsilon^2}{4}
+\frac{\epsilon K_1(\epsilon)}{K_0(\epsilon)}
\right]^{-1}.
\label{DCT}
\end{align}
In Fig.~\ref{fig:thermal_sd_es}, we plot $D^{\rm F}_{\rm T}$ as a function of 
the particle size $\delta$ by the solid line, and  $D^{\rm C}_{\rm T}$ as a function 
of $\epsilon$ by the dashed line for the whole size range. 
Their asymptotic behaviors are separately discussed below.

When we consider the total diffusion coefficient $D=D_{\rm T}+D_{\rm A}$,
we shall neglected the contribution from thermal fluctuations of force dipoles. 
These fluctuations can arise when force dipoles contain structural internal degrees of freedom.
However, such a contribution to the diffusion coefficient is small compared to $D_{\rm T}$
because it should be proportional to the product of $k_{\rm B}T$ and the concentration of 
force dipoles $c_0$.

\subsection{Free membranes} 

For the free membrane case, the total diffusion coefficient is given by 
$D^{\rm F}=D_{\rm T}^{\rm F}+D_{\rm A}^{\rm F}$, where the active non-thermal 
contribution $D_{\rm A}^{\rm F}$ was discussed in the previous section.
Using Eqs.~(\ref{DFT}) and (\ref{eq:diffusion_sd_small}) in the limit of $\delta\ll1$, we 
asymptotically have~\cite{Saf75,Saf76} 
\begin{align}
D^{\rm F}&\approx
\frac{k_{\rm B}T}{4\pi\eta_{\rm m}}\left(\ln\frac{2}{\nu\ell_{\rm c}}
-\gamma\right)
+\frac{Sc_0}{32\pi\eta_{\rm m}^2}\ln\frac{L}{\ell_{\rm c}},
\label{eq:total_sd_small}
\end{align}
where both contributions are proportional to $\ln (1/\ell_{\rm c})$.

For $\delta \gg 1$, on the other hand, we obtain from 
Eqs.~(\ref{DFT}) and (\ref{eq:diffusion_sd_large})~\cite{Hug81}
\begin{align}
D^{\rm F}&\approx
\frac{k_{\rm B}T}{16\eta_{\rm s}}\frac{1}{\ell_{\rm c}}
+\frac{5Sc_0}{256\pi\eta_{\rm s}^2}\frac{1}{\ell_{\rm c}^2}.
\label{eq:total_sd_large}
\end{align}
Since the $\ell_{\rm c}$-dependencies in Eq.~(\ref{eq:total_sd_large}) are different
between the thermal and non-thermal contributions, we can introduce a new crossover 
length defined by  
\begin{align}
\ell^{*}=\frac{5Sc_0}{16\pi k_{\rm B}T\eta_{\rm s}}. 
\label{eq:crossover_sd}
\end{align}
This length scale characterizes a crossover from the $1/\ell_{\rm c}^2$-dependence 
to $1/\ell_{\rm c}$-dependence. 
When $\ell_{\rm c} \ll \ell^{*}$ (but still $\nu^{-1} \ll \ell_{\rm c}$), the non-thermal 
contribution dominates over the thermal one, while in the opposite limit of 
$\ell_{\rm c} \gg \ell^{*}$, the thermal contribution is of primary importance.

\subsection{Confined membranes} 

In the case of confined membranes, the total diffusion coefficient now becomes 
$D^{\rm C}=D_{\rm T}^{\rm C} + D_{\rm A}^{\rm C}$. 
In the limit of $\epsilon \ll 1$, we have from Eqs.~(\ref{DCT}) and 
(\ref{eq:diffusion_es_small})~\cite{Evans88,Ramachandran10}
\begin{align}
D^{\rm C}&\approx
\frac{k_{\rm B}T}{4\pi\eta_{\rm m}}\left(\ln\frac{2}{\kappa\ell_{\rm c}}
-\gamma\right)
+\frac{Sc_0}{32\pi\eta_{\rm m}^2}\ln\frac{L}{\ell_{\rm c}},
\label{eq:total_es_small}
\end{align}
where both contributions exhibit a logarithmic dependence on $\ell_{\rm c}$
as in the free membrane case.

In the opposite limit of $\epsilon\gg1$, we find from Eqs.~(\ref{DCT}) and 
(\ref{eq:diffusion_es_large})~\cite{Evans88,Ramachandran10}
\begin{align}
D^{\rm C}&\approx
\frac{k_{\rm B}T}{2\pi\eta_{\rm s}}\frac{h}{\ell_{\rm c}^2}
+\frac{Sc_0}{16\pi\eta_{\rm s}^2}\frac{h^2}{\ell_{\rm c}^4}.
\label{eq:total_es_large}
\end{align}
Similar to the free membrane case, we can consider another characteristic length 
defined by
\begin{align}
\ell^{**}=\left(\frac{Sc_0h}{8k_{\rm B}T\eta_{\rm s}}\right)^{1/2}.
\label{eq:crossover_es}
\end{align}
This length scale characterizes a crossover from the $1/\ell_{\rm c}^4$-dependence 
to $1/\ell_{\rm c}^2$-dependence. 
We note that $\ell^{**}$ is essentially the geometric mean of $\ell^*$ and $h$. 
Numerical estimates of these two characteristic length scales will be discussed 
in Sec.~\ref{sec:discussion}.

\section{Drift velocity}
\label{sec:drift}

\subsection{Free membranes} 

\begin{figure}[tbh]
\begin{center}
\includegraphics[scale=0.35]{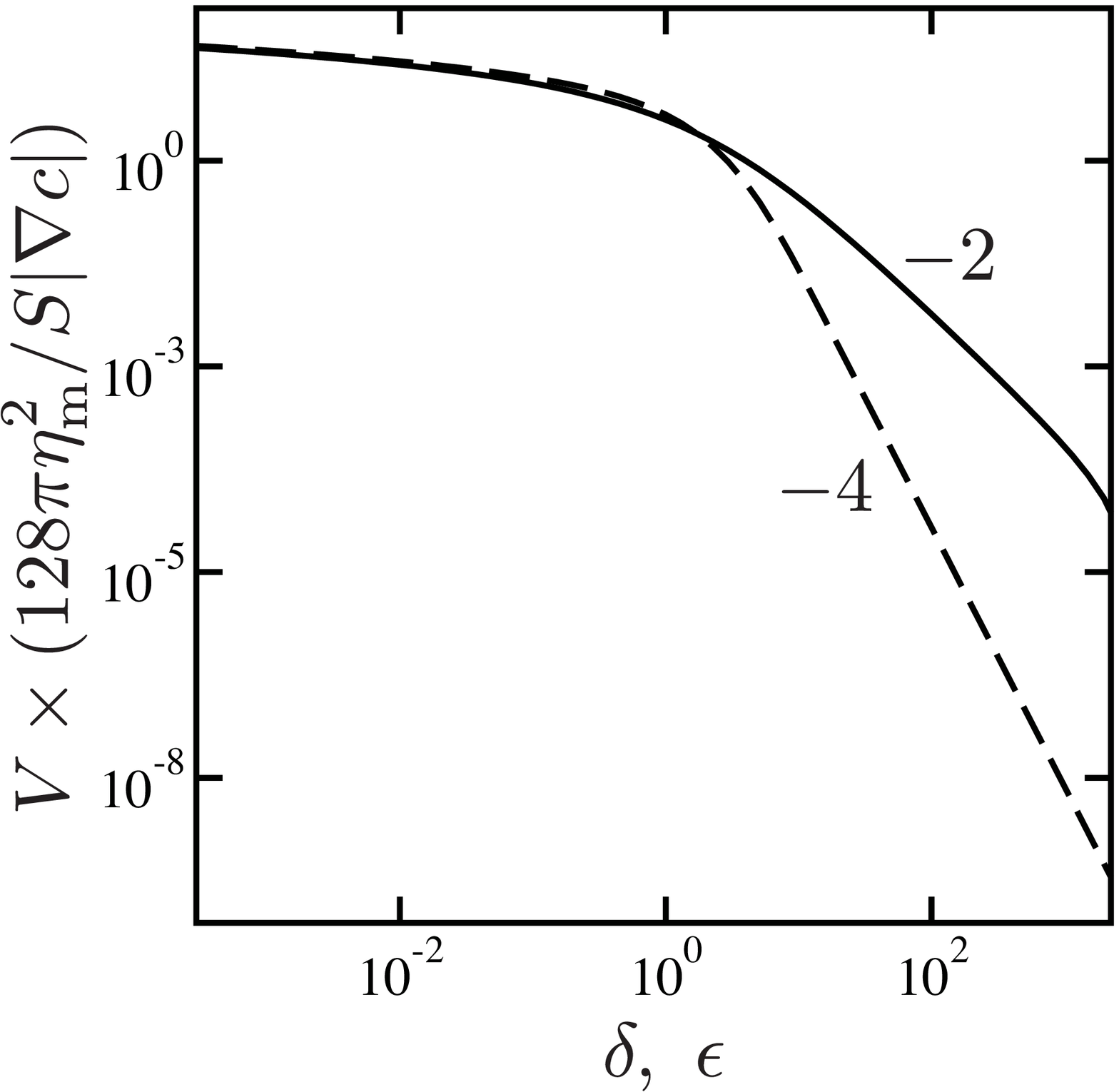}
\end{center}
\caption{
The plot of the scaled drift velocity $V$ as a function 
of the scaled cutoff length $\delta=\nu\ell_{\rm c}$ and $\epsilon =\kappa\ell_{\rm c}$ 
for the free membrane case [solid line, see Eq.~(\ref{eq:drift_sd})] and the confined 
membrane case [dashed line, see Eq.~(\ref{eq:drift_es})], respectively.
Here $V$ is scaled by $S  \vert \nabla c  \vert/(128\pi\eta_{\rm m}^2)$. 
The numbers in this plot indicate the slope of the curves and represent the powers 
of the algebraic dependencies.
}
\label{fig:drift_sd_es}
\end{figure}

In this section, we calculate the drift velocity $V$ of a passive particle due to 
a concentration gradient of active force dipoles. 
For the free membrane case, we substitute Eq.~(\ref{eq:mobility_sd}) into 
Eq.~(\ref{eq:drift}) and obtain 
\begin{align}
V^{\rm F}= & -\frac{S  \vert \nabla c  \vert}{16\pi^2\eta_{\rm m}^2}
\int_{\delta}^\infty {\rm d}^2z\,
\Omega_{\beta\beta^\prime\gamma\gamma^\prime} 
\nonumber \\
& \times \hat{n}_\alpha
\frac{\partial^2 g^{\rm F}_{\alpha\beta}(\mathbf{z})}{\partial z_\gamma\partial z_\delta}\frac{\partial g^{\rm F}_{\delta\beta^\prime}(\mathbf{z})}{\partial z_{\gamma^\prime}}
(\mathbf{z}\cdot\hat{\mathbf{n}}),
\label{eq:drift_sd}
\end{align}
where $\delta =\nu \ell_{\rm c}$ and 
$g^{{\rm F}}_{\alpha\beta}(\mathbf{z})=4\pi\eta_{\rm m}G^{\rm F}_{\alpha\beta}$ 
as before. 
Performing the numerical integration of Eq.~(\ref{eq:drift_sd}), we plot in 
Fig.~\ref{fig:drift_sd_es} the drift velocity $V^{\rm F}$ as a function of $\delta$  
by the solid line.
Similar to the active diffusion coefficient $D^{{\rm F}}_{\rm A}$, the drift velocity 
$V^{\rm F}$ depends weakly on the particle size at small scales, while it exhibits a 
stronger size dependence at large scales. 
Such a crossover also occurs around $\delta\approx1$.

We next discuss the asymptotic behaviors of $V^{\rm F}$ for small and large 
$\delta$ values.
With the use of Eqs.~(\ref{eq:mobility_sd_small}) and (\ref{eq:mobility_sd_large}), 
we show in the Appendix B that the asymptotic behaviors of $V$ for $\delta \ll 1$ 
and $\delta \gg 1$ are 
\begin{align}
V^{\rm F}&\approx\frac{S  \vert \nabla c  \vert}{32\pi\eta_{\rm m}^2}
\ln\frac{L}{\ell_{\rm c}},
\label{eq:drift_sd_small}
\end{align}
and
\begin{align}
V^{\rm F}&\approx\frac{13S  \vert \nabla c  \vert}{256\pi\eta_{\rm s}^2}\frac{1}{\ell_{\rm c}^2},
\label{eq:drift_sd_large}
\end{align}
respectively, where we choose $L \approx 1.85 \nu^{-1}$.
Note that Eq.~(\ref{eq:drift_sd_small}) was previously derived in Ref.~\cite{Mik15} for a 
2D membrane, while Eq.~(\ref{eq:drift_sd_large}) is a new result. 
As we see in Eqs.~(\ref{eq:drift_sd_small}) and (\ref{eq:drift_sd_large}), 
there is a crossover from a logarithmic to an algebraic dependence with a power of 
$-2$ when $\delta$ is increased.
These behaviors are consistent with the numerical plot in Fig.~\ref{fig:drift_sd_es} for 
the free membrane case.

\subsection{Confined membranes}

Finally we calculate the drift velocity for the confined membrane case.
Substituting Eq.~(\ref{eq:mobility_es}) into Eq.~(\ref{eq:drift}), we now obtain 
\begin{align}
V^{\rm C}=&-\frac{S  \vert \nabla c \vert}{16\pi^2\eta_{\rm m}^2}
\int_\epsilon^\infty {\rm d}^2w\,
\Omega_{\beta\beta^\prime\gamma\gamma^\prime}
\nonumber \\
& \times \hat{n}_\alpha
\frac{\partial^2 g^{\rm C}_{\alpha\beta}(\mathbf{w})}{\partial w_\gamma\partial w_\delta}\frac{\partial g^{\rm C}_{\delta\beta^\prime}(\mathbf{w})}{\partial w_{\gamma^\prime}}
(\mathbf{w}\cdot\hat{\mathbf{n}}),
\label{eq:drift_es}
\end{align}
where $\epsilon=\kappa\ell_{\rm c}$ and 
$g^{{\rm C}}_{\alpha\beta}(\mathbf{w})=4\pi\eta_{\rm m}G^{\rm C}_{\alpha\beta}$
as before. 
In Fig.~\ref{fig:drift_sd_es}, we present numerically calculated $V^{\rm C}$ as a 
function of $\epsilon$ by the dashed line.
As $\epsilon$ is increased, we see a crossover from a logarithmic to an algebraic dependence, 
although $V^{\rm C}$ decays faster than $V^{\rm F}$ at large scales.

The asymptotic behaviors of $V^{\rm C}$ for small and large $\epsilon$ values can be 
discussed similarly.
Using Eqs.~(\ref{eq:mobility_es_small}) and (\ref{eq:mobility_es_large}),
we obtain in the Appendix B the asymptotic expressions of $V^{\rm C}$ for 
$\epsilon \ll 1$ and $\epsilon \gg 1$ as
\begin{align}
V^{\rm C}&\approx\frac{S  \vert \nabla c  \vert}{32\pi\eta_{\rm m}^2}
\ln\frac{L}{\ell_{\rm c}},
\label{eq:drift_es_small}
\end{align}
and
\begin{align}
V^{\rm C}&\approx\frac{3S  \vert \nabla c  \vert}{16\pi\eta_{\rm s}^2}
\frac{h^2}{\ell_{\rm c}^4},
\label{eq:drift_es_large}
\end{align}
respectively, and we choose $L \approx 3.05 \kappa^{-1}$ to coincide with the numerical 
integration. 
We note that Eqs.~(\ref{eq:drift_sd_small}) and (\ref{eq:drift_es_small}) are identical and  
depend only on $\eta_{\rm m}$ for small sizes~\cite{Mik15}.

From Fig.~\ref{fig:drift_sd_es} and Eqs.~(\ref{eq:drift_sd_small}), (\ref{eq:drift_sd_large}), (\ref{eq:drift_es_small}) and (\ref{eq:drift_es_large}), we see that the drift velocity $V$ 
is always positive.
This means that passive particles move toward higher concentrations of active proteins,
and a chemotaxis-like drift takes place in the presence of protein concentration 
gradients~\cite{Mik15,Kap16,Koy16}.
The dominant viscosity dependence of $V$ switches from $\eta_{\rm m}$ to $\eta_{\rm s}$ 
as the particle size exceeds the corresponding hydrodynamic screening length, namely, 
$\nu^{-1}$ or $\kappa^{-1}$.

\section{Summary and Discussion}
\label{sec:discussion}

\begin{table*}
\caption{\label{tab:table3}
Summary of the asymptotic dependencies of the thermal diffusion coefficient $D_{\rm T}$, 
the active diffusion coefficient $D_{\rm A}$, and the drift velocity $V$ on 
the passive particle size $\ell_{\rm c}$. 
The numbers after the asymptotic expressions correspond to the equation numbers in this
paper.}
\begin{ruledtabular}
\begin{tabular}{ccccc}
cases &limits&$D_{\rm T}$&$D_{\rm A}$&$V$\\ \hline
free membrane&$\nu\ell_{\rm c}\ll1$&$\ln(1/\ell_{\rm c})$~~~(\ref{eq:total_sd_small})&$\ln(1/\ell_{\rm c})$~~~(\ref{eq:diffusion_sd_small})&$\ln(1/\ell_{\rm c})$~~~(\ref{eq:drift_sd_small})\\
($hk\gg1$)&$\nu\ell_{\rm c}\gg1$&$1/\ell_{\rm c}$~~~(\ref{eq:total_sd_large})&$1/\ell_{\rm c}^2$~~~(\ref{eq:diffusion_sd_large})&$1/\ell_{\rm c}^2$~~~(\ref{eq:drift_sd_large})\\ \hline
confined membrane &$\kappa\ell_{\rm c}\ll1$&$\ln(1/\ell_{\rm c})$~~~(\ref{eq:total_es_small})&$\ln(1/\ell_{\rm c})$~~~(\ref{eq:diffusion_es_small})&$\ln(1/\ell_{\rm c})$~~~(\ref{eq:drift_es_small})\\
($hk\ll1$)&$\kappa\ell_{\rm c}\gg1$&$1/\ell_{\rm c}^2$~~~(\ref{eq:total_es_large})&$1/\ell_{\rm c}^4$~~~(\ref{eq:diffusion_es_large})&$1/\ell_{\rm c}^4$~~~(\ref{eq:drift_es_large}) 
\end{tabular}
\end{ruledtabular}
\label{limits}
\end{table*}

In this paper, we have investigated lateral diffusion induced by active force dipoles embedded 
in a biomembrane.
In particular, we have calculated the active diffusion coefficient and the drift velocity for 
the free and the confined membrane cases by taking into account the hydrodynamic coupling 
between the membrane and the surrounding bulk solvent. 
The force dipole model in Refs.~\cite{Mik15,Kap16} and the general membrane mobility tensors 
obtained in Refs.~\cite{Inaura08,Ram11,Ram11b,Komura12} have been employed in our work.
When the size of a passive diffusing particle is small, the active diffusion coefficients for the 
free and the confined membranes represent the same logarithmic size dependence, as shown 
in Eqs.~(\ref{eq:diffusion_sd_small}) and (\ref{eq:diffusion_es_small}), respectively~\cite{Mik15}.
In the opposite large size limit, we find algebraic dependencies with powers $-2$ and 
$-4$ for the two cases, as given by Eqs.~(\ref{eq:diffusion_sd_large}) and 
(\ref{eq:diffusion_es_large}), respectively.
These are the important outcomes of this paper and are also summarized in Table~\ref{limits}
together with other asymptotic expressions.

In our work, we have assumed that the total diffusion coefficient is provided by the sum of 
thermal and non-thermal contributions. 
For small particle sizes, we have shown that both the total $D^{\rm F}$ and $D^{\rm C}$ exhibit 
a logarithmic size dependence~\cite{Mik15}, whereas different contributions have different size 
dependencies for large particle sizes. 
From this result, we have obtained two characteristic length scales that describe the 
crossover from non-thermal to thermal behaviors when the particle size is larger than
the hydrodynamic screening length.  
The drift velocity in the presence of a concentration gradient of active proteins exhibits
the same size dependencies as the active diffusion coefficient for the two membrane 
geometries.

Here we give some numerical estimates of the obtained crossover length scales. 
Using typical values such as 
$k_{\rm B}T\approx 4\times10^{-21}$\,J, 
$\eta_{\rm s}\approx10^{-3}$\,Pa$\cdot$s, 
$h\approx10^{-9}$\,m, 
$S\approx 10^{-42}$\,J$^2\cdot$s, and  
$c_0 \approx10^{14}$\,m$^{-2}$~\cite{Mik15}, we obtain 
$\ell^*\approx 2\times10^{-6}$\,m 
[see Eq.~(\ref{eq:crossover_sd})] and 
$\ell^{**} \approx 6\times10^{-8}$\,m 
[see Eq.~(\ref{eq:crossover_es})]. 
On the other hand, the SD and the ES screening lengths are typically 
$\nu^{-1}\approx5 \times 10^{-7}$\,m and 
$\kappa^{-1}\approx2\times10^{-8}$\,m, 
respectively~\cite{Saf75,Saf76,Evans88,Ramachandran10}.
Hence $\ell^*$ and $\ell^{**}$ are typically larger than $\nu^{-1}$ and $\kappa^{-1}$, 
respectively.
Moreover, the values of $S$ and $c_0$ can vary significantly in one membrane 
to another as pointed out in Ref.~\cite{Mik15}.
For example, when active proteins are confined in raft domains~\cite{Sim97,Kom14,Lin10}, 
the 2D concentration $c_0$ can be much larger. 
When, for example, $c_0 \approx10^{15}$\,m$^{-2}$ (while $S$ is the same as above)~\cite{Koy16}, 
the crossover length can be estimated as 
$\ell^*\approx 2\times10^{-5}$\,m 
and 
$\ell^{**} \approx 2\times10^{-7}$\,m.
If $\ell^{*}$ and $\ell^{**}$ are much larger than the screening lengths $\nu^{-1}$ and 
$\kappa^{-1}$, respectively, as in this case, the three different scaling regimes of 
the total diffusion coefficient are expected as the particle size is increased, i.e., 
$\ln(1/\ell_{\rm c}) \to 1/\ell_{\rm c}^2 \to1/\ell_{\rm c}$ 
for the free membrane case, 
and $\ln(1/\ell_{\rm c}) \to 1/\ell_{\rm c}^4 \to 1/\ell_{\rm c}^2$ for the confined 
membrane case.

Momentum in a membrane is conserved over distances smaller than the hydrodynamic 
screening length (either $\nu^{-1}$ or $\kappa^{-1}$), whereas it leaks to the surrounding 
fluid beyond that length scale~\cite{Dia09,Opp09,Opp10}.
Within a membrane, the velocity decays as $\ln(1/r)$ at short distances, 
as shown in Eqs.~(\ref{eq:mobility_sd_small}) and (\ref{eq:mobility_es_small}), 
due to the momentum conservation in 2D.
These 2D behaviors also lead to the logarithmic dependence of the active diffusion 
coefficients in Eqs.~(\ref{eq:diffusion_sd_small}) and (\ref{eq:diffusion_es_small}).
For the free membrane case, the velocity decays as $1/r$ at large scales as shown in 
Eq.~(\ref{eq:mobility_sd_large}) due to the momentum conservation in the 3D bulk.
This behavior is reflected in the first term of Eq.~(\ref{eq:total_sd_large}) for the
thermal diffusion coefficient~\cite{Hug81}.
As shown in Eq.~(\ref{eq:mobility_es_large}), however, the velocity decays as 
$1/r^2$ at large scales for the confined membrane case.
This behavior essentially arises from the mass conservation in 2D while the total 
momentum is not conserved due to the presence of the walls which break the 
translational symmetry of the system~\cite{Dia09,Opp09,Opp10}. 
The corresponding contribution is the first term of Eq.~(\ref{eq:total_es_large}) for 
the thermal diffusion coefficient~\cite{Evans88,Ramachandran10}.

The active diffusion coefficient $D^{{\rm F}}_{\rm A}$ obtained in 
Eq.~(\ref{eq:diffusion_sd_large}) for the free membrane case essentially reflects 
the hydrodynamics of the surrounding bulk 3D solvent.
Hence our result can be compared with that in Ref.~\cite{Mik15} obtained 
for a purely 3D fluid system:
\begin{align}
D^{{\rm 3D}}_{\rm A}&\approx\frac{S c_0^{\rm 3D}}{60\pi\eta_{\rm s}^2}
\frac{1}{\ell_{\rm c}},
\label{eq:diffusion_sd_bulk}
\end{align}
which decays as $1/\ell_{\rm c}$ and is different from Eq.~(\ref{eq:diffusion_sd_large}).
In fact, such a difference arises from the different dimensions of the dipole concentrations, 
i.e., $c_0$ is the 2D concentration in our case, while $c_0^{\rm 3D}$ is the 3D 
concentration in Ref.~\cite{Mik15}.
A similar comparison can be also made for the drift velocity of free membranes 
in Eq.~(\ref{eq:drift_sd_large}) and that in Ref.~\cite{Mik15} for a 3D fluid system:
\begin{align}
V^{\rm 3D}&\approx\frac{S  \vert \nabla c^{\rm 3D}  \vert}{30\pi\eta_{\rm s}^2}
\frac{1}{\ell_{\rm c}}.
\end{align}
The same reason holds for the different $\ell_{\rm c}$-dependence.

At this stage, we also comment that both the active diffusion coefficient $D_{\rm A}$ 
and the drift velocity $V$ exhibit the same $\ell_{\rm c}$-dependence.
Although the integrands in Eqs.~(\ref{eq:coefficient}) and (\ref{eq:drift}) look apparently
different, their physical dimensions are identical because the first derivative of 
the mobility tensor in Eq.~(\ref{eq:coefficient}) corresponds to the product of the 
second derivative and $(\mathbf{r}\cdot\hat{\mathbf{n}})$ in Eq.~(\ref{eq:drift}).
This is the simple reason that they exhibit the same $\ell_{\rm c}$-dependence.
One can also easily confirm that $V$ is positive when we make use of the membrane 
mobility tensor, because the integrand in Eq.~(\ref{eq:drift}) is the product of the first 
and the second derivatives of the mobility tensor which have opposite signs.
This leads to $V>0$ indicating a chemotaxis-like drift as mentioned before.

In this work, we have assumed that active proteins generate forces only in the lateral directions.
On the other hand, actual active motors such as bacteriorhodopsin can also exert forces to the 
surrounding solvent~\cite{Man99,Ramaswamy00,Man01}.
Although we did not take into account such normal forces which induce membrane undulation, 
consideration of normal forces as well as lateral ones will provide us with a general understanding 
of active diffusion in biomembranes~\cite{Komura15}.

We have also assumed that the force dipoles are fixed in a membrane and 
are distributed homogeneously.
It would be interesting to consider the case when active proteins can also 
move laterally in the membrane and even interact with each other through a nematic-like 
interaction~\cite{Lau09}.
The full equation of motion now involves potential-of-mean-force interactions in the 
multi-particle diffusion equations that describe the combined motions of the passive 
particle and active proteins in the membrane.
Although the dynamics of the active protein concentration is essentially determined by 
a diffusion equation, it is a complicated problem because not only thermal diffusion but 
also active non-thermal diffusion should be taken into account.  
Our work is the first step toward such a full description of very rich biomembrane dynamics.

\begin{acknowledgments}

We thank A.\  S.\ Mikhailov and T.\ Kato for useful discussions.
S.K.\ and R.O.\ acknowledge support from the Grant-in-Aid for Scientific Research on
Innovative Areas ``\textit{Fluctuation and Structure}" (Grant No.\ 25103010) from the Ministry
of Education, Culture, Sports, Science, and Technology of Japan, and the 
Grant-in-Aid for Scientific Research (C) (Grant No.\ 15K05250)
from the Japan Society for the Promotion of Science (JSPS).
\end{acknowledgments}

\appendix
\section{Derivation of Eqs.~(\ref{eq:diffusion_sd_large}) and (\ref{eq:diffusion_es_large})}
\label{app:diffusion}

Since Eqs.~(\ref{eq:diffusion_sd_small}) and (\ref{eq:diffusion_es_small})
have been obtained in Ref.~\cite{Mik15}, we show here the derivation of 
Eqs.~(\ref{eq:diffusion_sd_large}) and and (\ref{eq:diffusion_es_large}).
Substituting Eq.~(\ref{eq:mobility_sd_large}) into Eq.~(\ref{eq:diffusion_sd}),
we get 
\begin{align}
D^{\rm F}_{\rm A}&=\frac{Sc_0}{8\pi^2\eta_{\rm m}^2}
\int_\delta^\infty{\rm d}^2z\,
\Omega_{\beta\beta^\prime\gamma\gamma^\prime}\nonumber\\
&\times
\frac{\partial}{\partial z_\gamma}\left(\frac{\hat{z}_\alpha \hat{z}_\beta}{z}\right)
\frac{\partial}{\partial z_{\gamma^\prime}}\left(\frac{\hat{z}_\alpha \hat{z}_{\beta^\prime}}{z}\right),
\label{app:sd_large}
\end{align}
where $\mathbf{z}=\nu\mathbf{r}$.
Since 
\begin{align}
\frac{\partial}{\partial z_\gamma}\left(\frac{\hat{z}_\alpha \hat{z}_\beta}{z}\right)=&
\frac{1}{z^3}(\delta_{\alpha\gamma}z_\beta+\delta_{\beta\gamma}z_\alpha)
-\frac{3}{z^5}z_\alpha z_\beta z_\gamma,
\end{align} 
the integrand in Eq.~(\ref{app:sd_large}) becomes 
\begin{align}
&\frac{\partial}{\partial z_\gamma}\left(\frac{\hat{z}_\alpha \hat{z}_\beta}{z}\right)
\frac{\partial}{\partial z_{\gamma^\prime}}\left(\frac{\hat{z}_\alpha \hat{z}_{\beta^\prime}}{z}\right)
=\frac{1}{z^4}\delta_{\beta\gamma}\delta_{\beta^\prime\gamma^\prime}
\nonumber\\
& +\frac{1}{z^6}[\delta_{\gamma\gamma^\prime}z_\beta z_{\beta^\prime}-2(\delta_{\beta\gamma}z_{\beta^\prime}z_{\gamma^\prime}+\delta_{\beta^\prime\gamma^\prime} z_{\beta}z_{\gamma})]\nonumber\\
&+\frac{3}{z^8}z_\beta z_{\beta^\prime}z_\gamma z_{\gamma^\prime}.
\end{align}
By operating $\Omega_{\beta\beta^\prime\gamma\gamma^\prime}$, we have 
\begin{align}
\Omega_{\beta\beta^\prime\gamma\gamma^\prime}
\frac{\partial}{\partial z_\gamma}\left(\frac{\hat{z}_\alpha \hat{z}_\beta}{z}\right)
\frac{\partial}{\partial z_{\gamma^\prime}}\left(\frac{\hat{z}_\alpha \hat{z}_{\beta^\prime}}{z}\right)=
\frac{5}{8z^4}.
\end{align}
After the integration, we obtain Eq.~(\ref{eq:diffusion_sd_large}).

Similarly, we substitute Eq.~(\ref{eq:mobility_es_large}) into Eq.~(\ref{eq:diffusion_es}) and obtain 
\begin{align}
&D^{\rm C}_{\rm A}=\frac{Sc_0}{8\pi^2\eta_{\rm m}^2}
\int_\epsilon^\infty{\rm d}^2w\,
\Omega_{\beta\beta^\prime\gamma\gamma^\prime}
\nonumber \\
& \times
\frac{\partial}{\partial w_\gamma}\left(\frac{\delta_{\alpha\beta}-2\hat{w}_\alpha\hat{w}_\beta}{w^2}\right)
\frac{\partial}{\partial w_{\gamma^\prime}}\left(\frac{\delta_{\alpha\beta^\prime}-2\hat{w}_\alpha\hat{w}_{\beta^\prime}}{w^2}\right),
\label{app:es_small}
\end{align}
where $\mathbf{w}=\kappa\mathbf{r}$.
Since 
\begin{align}
&\frac{\partial}{\partial w_\gamma}\left(\frac{\delta_{\alpha\beta}-2\hat{w}_\alpha\hat{w}_\beta}{w^2}\right)\nonumber\\
&=-\frac{2}{w^4}(\delta_{\alpha\beta}w_\gamma+\delta_{\beta\gamma}w_\alpha+\delta_{\alpha\gamma}w_\beta)
+\frac{8}{w^6}w_\alpha w_\beta w_\gamma,
\end{align}
we obtain 
\begin{align}
&\frac{\partial}{\partial w_\gamma}\left(\frac{\delta_{\alpha\beta}-2\hat{w}_\alpha\hat{w}_\beta}{w^2}\right)
\frac{\partial}{\partial w_{\gamma^\prime}}\left(\frac{\delta_{\alpha\beta^\prime}-2\hat{w}_\alpha\hat{w}_{\beta^\prime}}{w^2}\right) \nonumber\\
& =\frac{4}{w^6}
\delta_{\beta\gamma}\delta_{\beta^\prime\gamma^\prime}
+\frac{4}{w^8}[
\delta_{\beta\beta^\prime}w_\gamma w_{\gamma^\prime}+
\delta_{\beta^\prime\gamma}w_\beta w_{\gamma^\prime}
+\delta_{\beta\gamma^\prime}w_{\beta^\prime} w_{\gamma}\nonumber \\
&+
\delta_{\gamma\gamma^\prime}w_\beta w_{\beta^\prime}
-2(\delta_{\beta\gamma}w_{\beta^\prime} w_{\gamma^\prime}+\delta_{\beta^\prime\gamma^\prime}w_\beta w_\gamma)
].
\end{align}
By operating $\Omega_{\beta\beta^\prime\gamma\gamma^\prime}$, we have
\begin{align}
&\Omega_{\beta\beta^\prime\gamma\gamma^\prime}
\frac{\partial}{\partial w_\gamma}\left(\frac{\delta_{\alpha\beta}-2\hat{w}_\alpha\hat{w}_\beta}{w^2}\right)
\frac{\partial}{\partial w_{\gamma^\prime}}\left(\frac{\delta_{\alpha\beta^\prime}-2\hat{w}_\alpha\hat{w}_{\beta^\prime}}{w^2}\right)\nonumber\\
&=\frac{4}{w^6}.
\end{align}
After the integration, we obtain Eq.~(\ref{eq:diffusion_es_large}).

\section{Derivation of Eqs.~(\ref{eq:drift_sd_large}) and (\ref{eq:drift_es_large})}
\label{app:drift}

In this Appendix, we show the derivation of Eqs.~(\ref{eq:drift_sd_large}) and (\ref{eq:drift_es_large}).
Substituting Eq.~(\ref{eq:mobility_sd_large}) into Eq.~(\ref{eq:drift_sd}), we obtain 
\begin{align}
V^{\rm F} =& -\frac{S  \vert \nabla c  \vert}{4\pi^2\eta_{\rm m}^2}
\int_{\delta}^\infty {\rm d}^2z\,
\Omega_{\beta\beta^\prime\gamma\gamma^\prime} 
\hat{n}_\alpha\nonumber\\
&\times\frac{\partial^2}{\partial z_\gamma\partial z_\delta}\left(
\frac{\hat{z}_\alpha\hat{z}_\beta}{z}
\right)
\frac{\partial}{\partial z_{\gamma^\prime}}\left(
\frac{\hat{z}_\delta \hat{z}_{\beta^\prime}}{z}
\right)
(\mathbf{z}\cdot\hat{\mathbf{n}}).
\end{align}
In the above, the derivatives are 
\begin{align}
&\frac{\partial^2}{\partial z_\gamma\partial z_\delta}\left(
\frac{\hat{z}_\alpha\hat{z}_\beta}{z}
\right)=\frac{1}{z^3}(
\delta_{\alpha\delta}\delta_{\beta\gamma}+\delta_{\alpha\gamma}\delta_{\beta\delta}
)\nonumber \\
&-\frac{3}{z^5}(\delta_{\alpha\delta}z_\beta z_\gamma+\delta_{\beta\delta}z_\alpha z_\gamma+\delta_{\alpha\gamma}z_\beta z_\delta+\delta_{\beta\gamma}z_\alpha z_\delta \nonumber \\
&+\delta_{\gamma\delta}z_\alpha z_\beta)
+\frac{15}{z^7}z_\alpha z_\beta z_\gamma z_\delta,
\end{align}
and
\begin{align}
&\frac{\partial^2}{\partial z_\gamma\partial z_\delta}\left(
\frac{\hat{z}_\alpha\hat{z}_\beta}{z}
\right)
\frac{\partial}{\partial z_{\gamma^\prime}}\left(
\frac{\hat{z}_\delta \hat{z}_{\beta^\prime}}{z}
\right) \nonumber \\
&=-\frac{1}{z^6}[
2\delta_{\beta^\prime\gamma^\prime}(\delta_{\alpha\gamma}z_\beta+\delta_{\beta\gamma}z_\alpha)
-(\delta_{\alpha\gamma^\prime}\delta_{\beta\gamma}+
\delta_{\alpha\gamma}\delta_{\beta\gamma^\prime})z_{\beta^\prime}
]\nonumber\\
&-\frac{3}{z^8}[(\delta_{\alpha\gamma^\prime}z_\beta z_\gamma+
\delta_{\beta\gamma^\prime}z_\alpha z_\gamma-
\delta_{\alpha\gamma}z_\beta z_{\gamma^\prime}-
\delta_{\beta\gamma}z_\alpha z_{\gamma^\prime}\nonumber \\
&+\delta_{\gamma\gamma^\prime}z_\alpha z_\beta) z_{\beta^\prime}
-2\delta_{\beta^\prime\gamma^\prime}z_\alpha z_\beta z_\gamma
]\nonumber \\
&-\frac{3}{z^{10}}z_\alpha z_\beta z_{\beta^\prime} z_\gamma z_{\gamma^\prime}.
\end{align}
By operating $\Omega_{\beta\beta^\prime\gamma\gamma^\prime}$, we have
\begin{align}
\Omega_{\beta\beta^\prime\gamma\gamma^\prime}
\frac{\partial^2}{\partial z_\gamma\partial z_\delta}\left(
\frac{\hat{z}_\alpha\hat{z}_\beta}{z}
\right)
\frac{\partial}{\partial z_{\gamma^\prime}}\left(
\frac{\hat{z}_\delta \hat{z}_{\beta^\prime}}{z}
\right)=-\frac{13z_\alpha}{8z^6}.
\end{align}
After the integration,  we obtain Eq.~(\ref{eq:drift_sd_large}).

Next we substitute Eq.~(\ref{eq:mobility_es_large}) into Eq.~(\ref{eq:drift_es}) and find
\begin{align}
V^{\rm C} =& -\frac{S  \vert \nabla c  \vert}{4\pi^2\eta_{\rm m}^2}
\int_{\epsilon}^\infty {\rm d}^2w\,
\Omega_{\beta\beta^\prime\gamma\gamma^\prime} 
\hat{n}_\alpha\nonumber \\
&\times \frac{\partial^2}{\partial w_\gamma\partial w_\delta}
\left(\frac{\delta_{\alpha\beta}-2\hat{w}_\alpha\hat{w}_\beta}{w^2}\right)
\nonumber \\
&\times
\frac{\partial}{\partial w_{\gamma^\prime}}
\left(\frac{\delta_{\delta\beta^{\prime}}-2\hat{w}_\delta\hat{w}_{\beta^{\prime}}}{w^2}\right)
(\mathbf{w}\cdot\hat{\mathbf{n}}).
\end{align}
Here the derivatives are 
\begin{align}
&\frac{\partial^2}{\partial w_\gamma\partial w_\delta}
\left(\frac{\delta_{\alpha\beta}-2\hat{w}_\alpha\hat{w}_\beta}{w^2}\right)\nonumber \\
&=-\frac{2}{w^4}(\delta_{\alpha\beta}\delta_{\gamma\delta}+\delta_{\alpha\gamma}\delta_{\beta\delta}+\delta_{\alpha\delta}\delta_{\beta\gamma})\nonumber\\
&+\frac{8}{w^6}(\delta_{\alpha\beta}w_\gamma w_\delta+\delta_{\beta\delta}w_\alpha w_\gamma+
\delta_{\alpha\delta}w_\beta w_\gamma+\delta_{\alpha\gamma}w_\beta w_\delta \nonumber \\
&+\delta_{\beta\gamma}w_\alpha w_\delta+\delta_{\gamma\delta}w_\alpha w_\beta)-\frac{48}{w^8}w_\alpha w_\beta w_\gamma w_\delta,
\end{align}
and
\begin{align}
&\frac{\partial^2}{\partial w_\gamma\partial w_\delta}
\left(\frac{\delta_{\alpha\beta}-2\hat{w}_\alpha\hat{w}_\beta}{w^2}\right)
\frac{\partial}{\partial w_{\gamma^\prime}}
\left(\frac{\delta_{\delta\beta^{\prime}}-2\hat{w}_\delta\hat{w}_{\beta^{\prime}}}{w^2}\right)  \nonumber \\
&=-\frac{4}{w^8}[
3\delta_{\beta^\prime\gamma^\prime}(\delta_{\alpha\beta}w_\gamma+\delta_{\alpha\gamma}w_\beta
+\delta_{\beta\gamma}w_\alpha)\nonumber \\
&-(\delta_{\alpha\beta}\delta_{\gamma\beta^\prime}+\delta_{\alpha\gamma}\delta_{\beta\beta^\prime}
+\delta_{\alpha\beta^\prime}\delta_{\beta\gamma})w_{\gamma^\prime}\nonumber \\
&-(\delta_{\alpha\beta}\delta_{\gamma\gamma^\prime}+\delta_{\alpha\gamma}\delta_{\beta\gamma^\prime}
+\delta_{\alpha\gamma^\prime}\delta_{\beta\gamma})w_{\beta^\prime}
]\nonumber\\
&+\frac{16}{w^{10}}[
(\delta_{\alpha\beta}w_{\beta^\prime} w_\gamma -\delta_{\beta\beta^\prime}w_\alpha w_\gamma -\delta_{\alpha\beta^\prime}w_\beta w_\gamma \nonumber \\
&+\delta_{\alpha\gamma}w_\beta w_{\beta^\prime}
+\delta_{\beta\gamma}w_\alpha w_{\beta^\prime}-\delta_{\beta^\prime\gamma}w_\alpha w_\beta
)w_{\gamma^\prime}\nonumber \\
&-(\delta_{\beta\gamma^\prime}w_\alpha w_\gamma +\delta_{\alpha\gamma^\prime}w_\beta w_\gamma 
+\delta_{\gamma\gamma^\prime}w_\alpha w_\beta )w_{\beta^\prime}\nonumber \\
&+3\delta_{\beta^\prime\gamma^\prime}w_\alpha w_\beta w_\gamma
].
\end{align}
By operating $\Omega_{\beta\beta^\prime\gamma\gamma^\prime}$, we find
\begin{align}
&\Omega_{\beta\beta^\prime\gamma\gamma^\prime}
\frac{\partial^2}{\partial w_\gamma\partial w_\delta}
\left(\frac{\delta_{\alpha\beta}-2\hat{w}_\alpha\hat{w}_\beta}{w^2}\right)\nonumber \\
&\times \frac{\partial}{\partial w_{\gamma^\prime}}
\left(\frac{\delta_{\delta\beta^{\prime}}-2\hat{w}_\delta\hat{w}_{\beta^{\prime}}}{w^2}\right)
=-\frac{12w_\alpha}{w^8}.
\end{align}
After the integration,  we obtain Eq.~(\ref{eq:drift_es_large}).


\end{document}